\documentclass[aps,pra,showpacs,amssymb,superscriptaddress,nofootinbib,notitlepage,11pt]{revtex4-1}

\usepackage{theorem,setspace,bm,enumerate,graphicx,epic,eepic,epsfig,amsmath,latexsym,amssymb,verbatim,revsymb,color,stmaryrd,fancyhdr,hyperref}

\newtheorem{definition}{Definition}

\newtheorem{lemma}[definition]{Lemma}

\newtheorem{theorem}[definition]{Theorem}
\newtheorem{corollary}[definition]{Corollary}

\def\squareforqed{\hbox{\rlap{$\sqcap$}$\sqcup$}}
\def\qed{\ifmmode\squareforqed\else{\unskip\nobreak\hfil
\penalty50\hskip1em\null\nobreak\hfil\squareforqed
\parfillskip=0pt\finalhyphendemerits=0\endgraf}\fi}
\def\endenv{\ifmmode\;\else{\unskip\nobreak\hfil
\penalty50\hskip1em\null\nobreak\hfil\;
\parfillskip=0pt\finalhyphendemerits=0\endgraf}\fi}

\newlength{\blank}
\newenvironment{proof}[1][{\hspace{-\blank}}]{{\noindent\textbf{Proof.   }}}{\hfill\qed\vskip 0.5\baselineskip}
\newenvironment{proofof}[1][{\hspace{-\blank}}]{{\noindent\textbf{Proof of }}}{\hfill\qed\vskip 0.5\baselineskip}

\mathchardef\ordinarycolon\mathcode`\:
\mathcode`\:=\string"8000
\def\vcentcolon{\mathrel{\mathop\ordinarycolon}}
\begingroup \catcode`\:=\active
  \lowercase{\endgroup
  \let :\vcentcolon
  }

\newcommand{\nc}{\newcommand}
\nc{\rnc}{\renewcommand}
\nc{\binomial}[2]{\genfrac{(}{)}{0pt}{}{#1}{#2}}
\nc{\lbar}[1]{\overline{#1}}
\nc{\bra}[1]{\langle#1|}
\nc{\ket}[1]{|#1\rangle}
\nc{\ketbra}[2]{|#1\rangle\!\langle#2|}
\nc{\braket}[2]{\langle#1|#2\rangle}
\nc{\Rank}{\operatorname{rank}\,}
\nc{\tr}{\operatorname{Tr}}

\nc{\cA}{{\cal A}}
\nc{\cB}{{\cal B}}
\nc{\cC}{{\cal C}}
\nc{\cD}{{\cal D}}
\nc{\cE}{{\cal E}}
\nc{\cF}{{\cal F}}
\nc{\cG}{{\cal G}}
\nc{\cH}{{\cal H}}
\nc{\cI}{{\cal I}}
\nc{\cJ}{{\cal J}}
\nc{\cK}{{\cal K}}
\nc{\cL}{{\cal L}}
\nc{\cM}{{\cal M}}
\nc{\cN}{{\cal N}}
\nc{\cO}{{\cal O}}
\nc{\cP}{{\cal P}}
\nc{\cR}{{\cal R}}
\nc{\cS}{{\cal S}}
\nc{\cT}{{\cal T}}
\nc{\cU}{{\cal U}}
\nc{\cX}{{\cal X}}
\nc{\cZ}{{\cal Z}}

\nc{\1}{{\openone}}

\def\e{\epsilon}

\def\r{\rho}
\def\s{\sigma}

\nc{\RR}{{{\mathbb R}}}
\nc{\CC}{{{\mathbb C}}}
\nc{\FF}{{{\mathbb F}}}
\nc{\NN}{{{\mathbb N}}}
\nc{\ZZ}{{{\mathbb Z}}}
\nc{\PP}{{{\mathbb P}}}
\nc{\QQ}{{{\mathbb Q}}}
\nc{\UU}{{{\mathbb U}}}
\nc{\EE}{{{\mathbb E}}}
\nc{\id}{{\operatorname{id}}}

\nc{\be}{\begin{equation}}
\nc{\ee}{\end{equation}}
\nc{\bea}{\begin{eqnarray}}
\nc{\eea}{\end{eqnarray}}

\nc{\LO}{\text{LO}}
\nc{\LOCC}{\text{LOCC}}
\nc{\cLOCC}{{\overline{\text{LOCC}}}}
\nc{\SEP}{\text{SEP}}
\nc{\PPT}{\text{PPT}}
\nc{\sep}{\text{sep}}
\nc{\twist}{\text{twist}}
\nc{\te}{\otimes}
\nc{\pro}[1]{\ket{#1}\!\bra{#1}}

\begin{document}
 \singlespacing
\title{Fundamental limitation on quantum broadcast networks}
\author{Stefan B{\"a}uml}
\email{stefan.bauml@lab.ntt.co.jp}
\affiliation{NTT Basic Research Laboratories, NTT Corporation, 3-1 Morinosato Wakamiya, Atsugi-shi, Kanagawa 243-0198, Japan}

\author{Koji Azuma}
\email{azuma.koji@lab.ntt.co.jp}
\affiliation{NTT Basic Research Laboratories, NTT Corporation, 3-1 Morinosato Wakamiya, Atsugi-shi, Kanagawa 243-0198, Japan}

\begin{abstract}
The ability to distribute entanglement over complex quantum networks is an important step towards a quantum internet. Recently, there has been significant theoretical effort, mainly focusing on the distribution of bipartite entanglement via a simple quantum network composed only of bipartite quantum channels. There are, however, a number of quantum information processing protocols based on multipartite rather than bipartite entanglement. Whereas multipartite entanglement can be distributed by means of a network of such bipartite channels, a more natural way is to use a more general network, that is, a quantum broadcast network including quantum broadcast channels. In this work, we present a general framework for deriving upper bounds on the rates at which GHZ states or multipartite private states can be distributed among a number of different parties over an arbitrary quantum broadcast network. Our upper bounds are written in terms of the multipartite squashed entanglement, corresponding to a generalisation of recently derived bounds [K.~Azuma {\it \it{et al.}}, Nat. Commun. {\bf 7}, 13523 (2016)]. We also discuss how lower bounds can be obtained by combining a generalisation of an aggregated quantum repeater protocol with graph theoretic concepts.
\end{abstract}
\date{\today}
\maketitle
\section{Introduction}

Distributing entanglement over long distances is an important prerequisite for the application of quantum protocols such as quantum key distribution (QKD) to real world communication problems \cite{PhysRevLett.67.661,bennett1992quantum,horodecki2009general}. The simplest way to do so is to create an entangled state locally and send part of it over a quantum channel. As the channel typically introduces noise and losses, it is usually necessary to send many copies of the state via the channel and then perform local operations and classical communication (LOCC) in order to distil the desired resource state \cite{bennett1996mixed,horodecki2009general}. The channel might however be too noisy or lossy to transmit entanglement at any feasible rate. For example, a typical quantum channel, such as an optical fibre, has an absorption rate that increases exponentially with the channel length. As a result, if the optical channel is longer than several hundred kilometres, entanglement distribution over the single channel would be inefficient in practice \cite{scarani2009security}.

On the other hand, a lot of theoretical progress has been made for this kind of point-to-point quantum communication.
In fact, recently, Takeoka {\it \it{et al.}} have shown \cite{takeoka2014fundamental,takeoka2014squashed} that the asymptotic rate at which a secret key or entanglement can be transmitted via many uses of a channel assisted by two-way classical communication is upper bounded by the squashed entanglement \cite{christandl2004squashed} of that channel. Besides, using the squashed entanglement, they have upper bounded the two-way assisted quantum/private capacities of the pure-loss channel and the thermal channel. Goodenough {\it \it{et al.}} \cite{goodenough2015assessing} have computed upper bounds on the squashed entanglement for several commonly used channels, such as phase-insensitive Gaussian bosonic channels \cite{weedbrook2012gaussian}. A different bound in terms of the relative entropy of entanglement \cite{vedral1997quantifying} of the channel has been provided \cite{2015arXiv151008863P,wilde2016converse}. This bound is tighter than the one based on the squashed entanglement for a number of teleportation-covariant channels \cite{bennett1996mixed}, succeeding in determining their two-way assisted private/quantum capacities. But, it is still an open question which of those bounds is tighter for general quantum channels. A strong converse for the bound based on the relative entropy of entanglement has been shown \cite{wilde2016converse}, meaning that the error rate quickly tends towards one if the rate exceeds the bound. Other strong converse bounds, in terms of the max relative entropy of entanglement \cite{datta2009min}, on the quantum and private capacities assisted by two-way classical communication have been provided by M{\"u}ller-Hermes {\it \it{et al.}} \cite{muller2016positivity} and Christandl {\it \it{et al.}} \cite{christandl2016relative}, respectively. It is also shown that there exist channels for which the max relative entropy of entanglement provides a bound significantly better than the squashed entanglement \cite{takeoka2014fundamental,takeoka2014squashed}.

Despite these results, the limitations of point-to-point entanglement transmission can be overcome by use of quantum repeaters \cite{briegel1998quantum}. In the repeater scenario, a sender (Alice) and a receiver (Bob) are connected by a chain of intermediate vertices and quantum channels. Entanglement between Alice and Bob is established by distributing entangled states between adjacent vertices and performing LOCC. A lot of effort has been put into optimising the performance and realisability of quantum repeaters (see \cite{sangouard2011quantum,abruzzo2013quantum,munro2015inside,azuma2015all} and references therein). The results of \cite{bauml2015limitations} provide a fundamental insight into which resources are required in quantum repeaters. Namely, it is shown that bound entanglement between adjacent vertices is, in some cases, insufficient to distribute privacy between Alice and Bob via a quantum repeater, suggesting that distillable entanglement might be necessary for distributing privacy over long distances.

While quantum repeaters make distribution of entanglement over arbitrarily large distances possible, a general quantum internet \cite{kimble2008quantum} will distribute entanglement in a more efficient manner over complex networks rather than just a linear chain of vertices. For example, entanglement between two parties on two different continents might be distributed using different undersea cables, depending on the traffic. In \cite{azuma2016fundamental}, the upper bound on the obtainable bipartite key/entanglement generation rate for point-to-point quantum communication \cite{takeoka2014fundamental,takeoka2014squashed} has been generalised to be applicable to an arbitrary network consisting of ancillary vertices and arbitrary quantum channels. Likewise, Pirandola generalises \cite{pirandola2016optimal} the upper bound given in \cite{2015arXiv151008863P} to arbitrary networks composed of teleportation-covariant channels, establishing a number of repeater-assisted capacities under various forms of system routing and extending classical results of network information theory to the quantum setting. In \cite{azuma2016aggregating}, a lower bound on the bipartite key/entanglement generation rate has also been presented. The lower bound is derived by introducing a so-called aggregated repeater protocol and applying a graph theoretic argument known as Menger's theorem \cite{menger1927allgemeinen}. However, this protocol considers only two-client cases.

More generally, there are many cryptographic problems involving more than two clients. For example, there is a scenario where a family of users needs a common cryptographic key, such that they can communicate openly among them but securely against external eavesdroppers. It has been shown that such a multipartite key can be obtained from a GHZ state or from a more general class of multipartite entangled states known as multipartite private states \cite{augusiak2009multipartite}. Another cryptographic protocol involving many clients is secret sharing, where two or more users have to come together in order to decrypt a message. It has been shown that this can be achieved using a GHZ state \cite{hillery1999quantum}. Other tasks requiring the distribution of multipartite entanglement in quantum networks include distributed quantum computation as well as the synchronisation of a network of atomic clocks, which can also be achieved with a GHZ state \cite{komar2014quantum}. A protocol to distribute graph states, which include GHZ states as special cases, in an arbitrary network of quantum channels has recently been introduced \cite{epping2016large,epping2016quantum}. An alternative way to distribute GHZ states in highly structured networks may be a so-called 2D quantum repeater protocol \cite{wallnofer20162d}. While it is possible to distribute multipartite entanglement in a network of single-sender-single-receiver channels, a more natural way to do so is to use quantum broadcast channels, i.e., channels with a single sender and multiple receivers. For instance, if we look at the conventional internet, it includes optical broadcast channels for the so-called last-mile service to the end user and wireless networks. As a quantum version of the internet would include similar elements \cite{frohlich2013quantum,hughes2013network}, it is worth going beyond the results of \cite{azuma2016fundamental}. The authors of \cite{seshadreesan2015bounds,laurenza2016general} have upper bounded the rates at which GHZ and multipartite private states can be distributed using a single broadcast channel. Further, Takeoka {\it et al.} \cite{takeoka2016unconstrained} have provided upper and lower bounds on the rate at which bipartite key and maximally entangled states can be distributed between a sender and many receivers in parallel via a pure-loss bosonic broadcast channel. The lower bound is achieved by a protocol based on quantum state merging \cite{horodecki2007quantum}. The bounds become tight in the limit of infinite average photon numbers.

In the present work, we present a general framework that allows us to derive upper bounds on the rates at which GHZ states or multipartite private states can be distributed among arbitrary families of users in parallel over a general quantum network including quantum broadcast channels, that is, over a quantum broadcast network. The upper bounds are written in terms of the multipartite squashed entanglement \cite{yang2009squashed,avis2008distributed}. Our results are obtained by combining the network approach of \cite{azuma2016fundamental} with the broadcast channel scenario presented in \cite{seshadreesan2015bounds}. As a result, our upper bound is a generalised version of those in \cite{azuma2016fundamental,seshadreesan2015bounds}. In addition, we discuss how the  aggregated repeater protocol presented in \cite{azuma2016aggregating} can be generalised to multi-user scenarios. In combination with a suitable generalisation of Menger's theorem, this can provide lower bounds on the achievable rates. The proposed generalisation of the aggregated repeater protocol to quantum broadcast networks differs from the state merging based protocol \cite{horodecki2007quantum} in that it can be used for the distribution not only of Bell or bipartite private states but also of GHZ or multipartite private states.

The intuition behind our main result, Theorem~\ref{EsqBoundMark}, is as follows. Any quantum internet protocol, finally producing multipartite entangled states among distant vertices, needs to utilise flows of quantum information via quantum broadcast channels. What kind of flow is generated depends on the protocol, which might be complex. Our strategy is to categorise such flows, depending on which flow could contribute to establishing which multipartite entangled state. This categorisation is specified by the partition ${\cal P}$ in Theorem~\ref{EsqBoundMark}. A partition ${\cal P}$ divides all the vertices in the network into classes. For a given partition we can look at the protocol as a way of supplying entanglement among the classes. How much entanglement can be distributed among the classes is upper bounded only by the capacities of the broadcast channels connecting the different classes, rather than those of all the given channels. This intuition leads to the inequality in Theorem~\ref{EsqBoundMark}. 

This paper is organised as follows. In section \ref{sec:pre} we introduce some concepts of multipartite entanglement we will use. In section \ref{sec:Notation} we describe the network architecture and protocols used in this work and introduce some notation. We then formulate our main results, i.e. upper bounds on the achievable rates, in section \ref{sec:upper}. Section \ref{sec:lower} contains our ideas regarding how the aggregated repeater protocol can be generalised. We conclude with section \ref{sec:concl}, where we discuss some open questions.

\section{Preliminaries}\label{sec:pre}
In this section, we briefly introduce some concepts and notation used in our paper. 
Multipartite entanglement---whose distribution among distant clients will be regarded as the goal of our protocols---is well known to have a structure richer than the bipartite one. For instance, in contrast to the bipartite case, there is no unique way to define a maximally entangled state. Indeed, there are different classes of `maximally' entangled states---such as GHZ states \cite{greenberger1989going} and W states---that cannot be converted into each other by means of LOCC, even in a probabilistic manner \cite{dur2000three}. 

One of the representative classes of multipartite entanglement is the family of GHZ states. In fact, GHZ states have been shown to be useful for achieving secret sharing, multi-party QKD and clock synchronisation. 
The $m$-qudit GHZ state is defined as
\be
\ket{\Phi_d}^{A_1\cdots A_m}=\frac{1}{\sqrt{d}}\sum_{i=0}^{d-1}\ket{i\cdots i}^{A_1\cdots A_m}
\ee
with an orthonormal basis $\{\ket i\}$.
In the case of $m=2$, the GHZ state is reduced to a bipartite maximally entangled state. Similarly to the bipartite case, if $m$ parties $A_1,A_2,\ldots,A_m$ share the GHZ state $\ket{\Phi_d}^{A_1\cdots A_m}$, they can obtain an $m$-partite secret key just by performing local projective measurements in the basis $\{\ket{i}\}$. However, the GHZ state $\ket{\Phi_d}^{A_1\cdots A_m}$ is not the only state from which we can distil an $m$-partite secret key. In fact, there is a larger class of states, called $m$\emph{-partite private states}, which can provide an $m$-partite secret key \cite{augusiak2009multipartite}. Such $m$-partite private states have been shown to be of the form
\begin{align}\label{pdit}
\gamma_d^{A_1\cdots A_m}&=U^\twist (\pro{\Phi_d}^{A'_1\cdots A'_m} \te\s^{A''_1\cdots A''_m} )U^{\twist\dagger}\\
&=\frac{1}{d}\sum_{k,i=0}^{d-1}\ketbra{i\cdots i}{k\cdots k}^{A'_1\cdots A'_m}\te U_{i}^{A''_1\cdots A''_m}\s^{A''_1\cdots A''_m}U_{k}^{A''_1\cdots A''_m\dagger}, 
\end{align}
where $A'_i$ and $A''_i$ are systems held by party $A_i$, $\s^{A''_1\cdots A''_m}$ is an arbitrary state and
\be
U^\twist=\sum_{i_1,\ldots,i_m=0}^{d-1}\pro{i_1\cdots i_m}^{A'_1\cdots A'_m}\te U_{i_1\cdots i_m}^{A''_1\cdots A''_m}
\ee
is a controlled unitary. This unitary is called twisting because it `twists' the entanglement present in the GHZ state into a more complex form also involving subsystems $A''_1\cdots A''_m$. The key is obtained from the state $\gamma_d^{A_1\cdots A_m}$ by performing projective measurements in the basis $\{\ket{i_1\cdots i_m}\}$ on subsystems $A'_1\cdots A'_m$, which are called the \emph{key part}. Its security is guaranteed if subsystems $A''_1\cdots A''_m$, which is called the {\it shield part}, are kept away from adversaries. 

While a private state with a large shield part is of limited practical use, it is interesting from a theoretical point of view. In fact, it has been shown that there exist \emph{bound entangled} states from which we cannot distil GHZ states even in the asymptotic setting but can obtain states arbitrarily close to a private state \cite{augusiak2009multipartite}, showing that GHZ-state distillability is not necessary for secret-key distillation.

The GHZ state could be distributed directly via a quantum broadcast channel \cite{yard2011quantum}. A quantum broadcast channel is a quantum channel $\cN:{x\to y_1\cdots y_{r}}$ that sends a subsystem $x$ of a sender to many receivers with respective outputs $y_1,y_2,\ldots,y_r$. Its idealised version is an isometry ${\cal I}^{x \to y_1\cdots y_r }=\sum_{i=0}^{d-1}\ket{i\cdots i}^{y_1\cdots y_r}\bra{i}^{x}$, which can be used to distribute the GHZ state as
\be
{\cal I}^{x \to y_1\cdots y_r } \ket{\Phi_d}^{x'x} =\ket{\Phi_d}^{x'y_1\cdots y_{r}}
\ee
in an ideal manner.

To evaluate multipartite entanglement, we use the multipartite squashed entanglement \cite{yang2009squashed,avis2008distributed}, which is defined as  
\be
E_{\rm sq}^{A_1:\cdots:A_m}(\r)=\inf_{\sigma:{\rm Tr}_E \sigma = \r } I(A_1:\cdots :A_m|E)_\sigma,
\ee
where the infimum is taken over extensions $\sigma^{A_1\cdots A_m E}$ with ${\rm Tr}_E(\sigma^{A_1\cdots A_m E})=\r^{A_1\cdots A_m}$ and the multipartite conditional mutual information is defined as 
\be
I(A_1:\cdots:A_m|E)=\sum_{i=1}^m H(A_i|E)-H(A_1\cdots A_m|E).
\ee
As in the bipartite case, the multipartite squashed entanglement is additive on tensor products and asymptotically continuous. It has also been shown \cite{seshadreesan2015bounds} that the multipartite squashed entanglement does not increase under groupings, i.e.
\be
E_{\rm sq}^{A_1:A_2:\cdots:A_m}(\r^{A_1A_2\cdots A_m})\ge E_{\rm sq}^{A_1A_2:\cdots:A_m}(\r^{A_1A_2\cdots A_m}).
\ee
Further we have the reduction property
\be\label{reduction}
E_{\rm sq}^{A_1:A_2:\cdots:A_m}(\s^{A_1}\te\r^{A_2\cdots A_m})=E_{\rm sq}^{A_2:\cdots:A_m}(\r^{A_2\cdots A_m}).
\ee
It has also been shown that
\be
E_{\rm sq}^{A_1:A_2:\cdots:A_m}(\pro{\Phi_d}^{A_1\cdots A_m})=m\log d
\ee
for GHZ state $\ket{\Phi_d}^{A_1\cdots A_m}$.
As the shield part of a private state can contain entanglement in addition to the key part, it holds 
\be
E_{\rm sq}^{A_1:A_2:\cdots:A_m}(\gamma_d^{A_1\cdots A_m})\ge m\log d
\ee
for private states $\gamma_d^{A_1\cdots A_m}$. 

\section{Quantum broadcast network}\label{sec:Notation}
In this section we briefly describe the concept of quantum broadcast networks and introduce some notation we will use. A quantum broadcast network can be associated with a directed hypergraph $H=(\cal V, E)$, where ${\cal V}$ is a set of vertices and ${\cal E}$ is a set of directed hyperedges.
 
The vertices of ${\cal V}$ represent quantum nodes which are allowed to use arbitrary LOCC among them. If a quantum state $\rho$ is shared by a set ${\cal V}' \subset {\cal V}$ of nodes, we write it as $\rho^{{\cal V'}}$ in what follows.
Besides, the quantum system of the state $\rho^{{\cal V'}}$ is denoted by ${\cal H}({\cal V'})$. 
In order to prove our results we will also need to introduce the concept of partitions. A ($k$-partite) partition ${\cal P=P}_1:\cdots:{\cal P}_k$ divides $\cal V$ into $k$ disjoint non-empty sets $\cP_1,\ldots,\cP_{k}$ of vertices such that $\cal V$ is the union of all the classes. That is, $\cP_i\neq\emptyset$ for any $i$, $\cP_i \cap \cP_j =\emptyset $ for $i\neq j$ and $\bigcup_{j=1}^k \cP_j={\cal V}$.

\begin{figure}
\includegraphics[width=\textwidth]{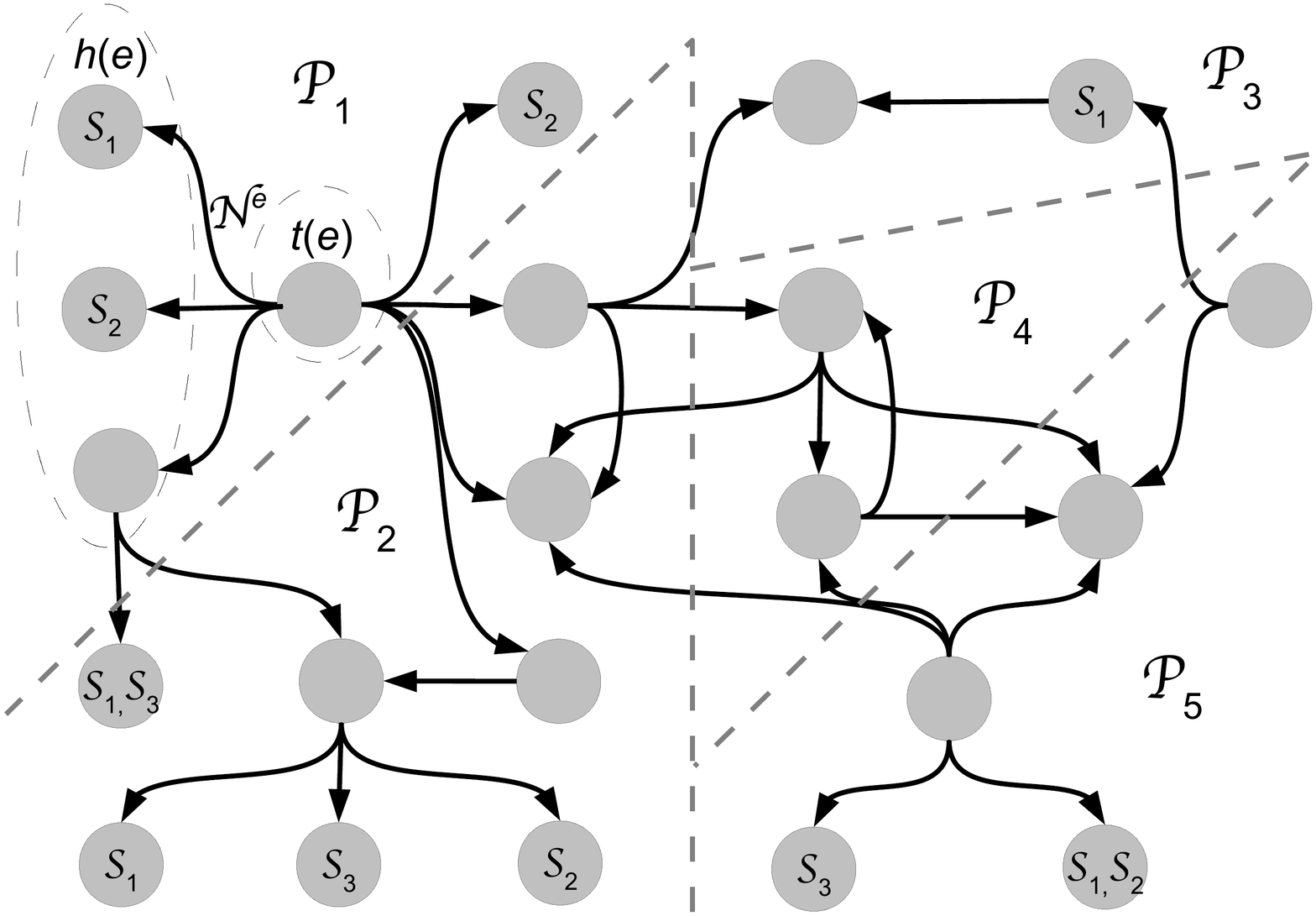}
	\caption{Example of a quantum broadcast network. The arrows correspond to quantum broadcast channels. The dots correspond to vertices. In this example, there are three families $\cS_1, \cS_2,\cS_3$ of clients. The vertices belonging to those families are labelled accordingly. The unlabelled vertices serve as repeater stations. Further, we have included an example of a partition $\cP=\cP_1:\cdots:\cP_5$ into five classes. The dashed lines correspond to boundaries between the classes. As an example, we show the head $h(e)$ and tails $t(e)$ for a broadcast channel $\cN^e$.}
	\label{fig1}
\end{figure}

In addition to LOCC, quantum nodes can use given quantum broadcast channels which are associated with directed hyperedges in the set $\cE$, respectively.  
In particular, such a hyperedge $e \in \cE $ [also described by $t(e) \to h(e)$ with the tail $t(e)$ and the heads $h(e)$] represents a quantum broadcast channel in the quantum network by ${\cal N}^e$, 
its tail $t(e) (\subset {\cal V})$ only with a single vertex indicates the quantum node holding the input of the channel ${\cal N}^e$ and its heads $h(e) (\subset {\cal V})$, perhaps with many vertices, mean the quantum nodes to receive the output systems of the channel ${\cal N}^e$.
See also figure \ref{fig1}. 

The goal of a protocol can be specified by $m$ sets ${\cal S}_1(\subset {\cal V}),\ldots,{\cal S}_m(\subset {\cal V})$ of clients, each ${\cal S}_i$ of which wishes to establish $|{\cal S}_i|$-partite entanglement among the clients in the nodes of ${\cal S}_i$. 
There might be cases where an intersection of these sets is not empty. We may also use ${\cal T}_m:=\bigcup_{j=1}^m {\cal S}_j$. All vertices in $\cal V$ that are not part of a family of clients may serve as repeater stations. We call the set of repeater stations ${\cal S}_0$, that is, ${\cal S}_0:={\cal V} \setminus {\cal T}_m$. 
Then, the goal is to supply the sets ${\cal S}_1,\ldots,{\cal S}_m$ of clients with multipartite private states $\gamma^{\cS_1}_{d^{(1)}},\ldots, \gamma^{\cS_m}_{d^{(m)}}$, respectively, in parallel. 
Hence, our target state is of the form
\be\label{target}
\tau_{\bm{d}}^{{\cal T}_m}=\bigotimes_{j=1}^{m}\gamma^{\cS_j}_{d^{(j)}},
\ee
where we have defined $\bm{d}=(d^{(1)},\ldots,d^{(m)})$. 
Note that the multipartite private state $\gamma^{\cS_j}_{d^{(j)}}$ also includes a $d^{(j)}$-dimensional GHZ state as a special case. Thus, our results hold irrespectively of whether $\gamma^{\cS_j}_{d^{(j)}}$ represents a GHZ state or a private state.

In order to obtain the target state, we need to use a quantum internet protocol. 
In general, a quantum internet protocol can be described as follows: Suppose that the initial state of the whole system $\cal V$ is described by ${\rho}_0^{{\cal V}}$. Then, the protocol begins with application of broadcast channel $\cN^{e_0}$ with $e_0\in {\cal E}$, where we define $\bm{k}_0=k_0=0$, followed by a round of (probabilistic) LOCC. 
The outcome $k_1$ of this LOCC, which appears with probability $p(k_1)$, may determine the broadcast channel $\cN^{e_{k_1}}$ with $e_{k_1}\in {\cal E}$ to be used next. After the use of channel $\cN^{e_{k_1}}$, another round of LOCC is performed, presenting an outcome $k_2$ with probability $p(k_2|k_1)$. Depending on previous outcomes $\bm{k}_2=(k_1,k_2)$, we may use quantum channel $\cN^{e_{\bm{k}_2}}$ with $e_{\bm{k}_2} \in {\cal E}$, followed by LOCC providing an outcome $k_3$ with probability $p(k_3|\bm{k}_2)$. Similar operations are continued until the final round, say round $l$. As a result, the protocol supplies quantum state $\rho_{\bm{k}_l}^{{\cal V}}$ with probability $p({\bm k}_l)=p(k_l|{\bm k}_{l-1}) \cdots p(k_3|\bm{k}_2) p(k_2|k_1) p(k_1)$, whose reduced density operator for nodes ${\cal T}_m$ should be close to our target state $\tau_{\bm{d}_{{\bm k}_l}}^{{\cal T}_m}$.


Finally, let us introduce a simple notation of the squashed entanglement of states and broadcast channels w.r.t. a given partition of the network:
Let ${\cal P}=\cP_1:\cdots:\cP_k$ be a partition of the set $\cal V$ of vertices. For a state $\r^{\cal V}$, we define 
\be\label{Esq1}
E^\cP_{\rm sq}(\r^{\cal V}):= E^{{\cal H}(\cP_1):\cdots:{\cal H}(\cP_k)}_{\rm sq}(\r^{\cal V}).
\ee
For a state $\r^{{\tilde {\cal V}}}$ on a subset ${\cal\tilde V}$ of ${\cal V}$ $(i.e., {\cal \tilde{V}} \subset {\cal V})$, we define
\be\label{Esq2}
E^\cP_{\rm sq}(\r^{ \cal  \tilde V}):=E^{{\cal H}(\cP_1\cap{\cal\tilde V}):\cdots:{\cal H}(\cP_k\cap{\cal \tilde V})}_{\rm sq}(\r^{\cal \tilde V}),
\ee
where, if $\cP_j \cap {\cal \tilde V}$ is empty, we strip ${\cal H}(\cP_j \cap {\cal \tilde V})$ from the partition in the right-hand side of this equation\footnote{That is, if $\cP_j \cap {\cal \tilde V}$ is empty, 
${\cal H}(\cP_1\cap{\cal\tilde V}):\cdots:{\cal H}(\cP_{j-1}\cap{\cal\tilde V}):{\cal H}(\cP_j\cap{\cal\tilde V}):{\cal H}(\cP_{j+1}\cap{\cal\tilde V} ):\cdots:{\cal H}(\cP_k\cap{\cal\tilde V})={\cal H}(\cP_1\cap{\cal\tilde V}):\cdots:{\cal H}(\cP_{j-1}\cap{\cal\tilde V}):{\cal H}(\cP_{j+1}\cap{\cal\tilde V} ):\cdots:{\cal H}(\cP_k\cap{\cal\tilde V}$). Note that we implicitly define $E_{\rm sq}^{\cal P}(\rho^{\tilde{\cal V}} )=0$ unless there are two or more different values of $j$ with ${\cal P}_j \cap{\cal \tilde{V}} \neq \emptyset $.}.
For a broadcast channel ${\cal N}^e$ with input system $X$, we define the squashed entanglement of the channel w.r.t. $\cP$ as
\be\label{def:Esq}
E^{\cP}_{{\rm sq}}\left(\cN^e\right)=\max_{\ket{\psi}^{RX}} E^{\cP}_{{\rm sq}}\left(\id^R\te\cN^{e}\left(\pro{\psi}^{RX}\right)\right),
\ee
where the maximisation is taken over all pure states $\ket{\psi}^{RX}$ that can be prepared at $t(e)$ locally. The maximisation can be restricted to pure states because the multipartite squashed entanglement is convex \cite{yang2009squashed}. 
Up to a normalisation factor $1/2$, equations (\ref{Esq1})-(\ref{def:Esq}) are equivalent to corresponding definitions\footnote{While in \cite{seshadreesan2015bounds} there is no explicit definition of the squashed entanglement of a broadcast channel, upper bounds as in Theorem 13 of \cite{seshadreesan2015bounds} can also be expressed in terms of the squashed entanglement of a channel as in equation (\ref{def:Esq}).} in \cite{seshadreesan2015bounds}. Equation (\ref{def:Esq}) reduces to the definition given in \cite{takeoka2014fundamental}, in the case of simple quantum channels that connect two vertices. For a given partition ${\cal P}$, it is convenient to define the set ${\cal E}_{\rm tri}^{\cal P}$ of edges $e\in {\cal E}$ whose tail and heads all belong to one set of $\cP$. In fact, if $e\in {\cal E}_{\rm tri}^{\cal P}$, the corresponding channel ${\cal N}^e$ is considered to be a local channel for the partition ${\cal P}$, represented by $E^{\cP}_{{\rm sq}}\left(\cN^e\right)=0$.

\section{Upper bounds}\label{sec:upper}

In this section, we present our main conclusions, that is, upper bounds on the sizes of GHZ states or multipartite private states obtainable by using a quantum internet protocol over a quantum broadcast network. We begin by showing that the obtainable squashed entanglement throughout a protocol is upper bounded by the initial squashed entanglement and the squashed entanglement of the used nontrivial channels for the partition ${\cal P}$.

\begin{theorem}\label{lemma:main}
For any adaptive $l$-round protocol resulting in $\r^{\cal V}_{{\bm k}_l}$ with probabilities $p({\bm k}_l)$ by using a quantum broadcast network associated with a directed hypergraph $H=(\cal V, E)$, it holds
\be
\langle E_{\rm sq}^{\cP}(\r^{\cal V}_{{\bm k}_l}) \rangle_{{\bm k}_l}\le E_{\rm sq}^{\cP}(\r_0^{{\cal V}})+\sum_{e\in {\cal E}\setminus {\cal E}_{\rm tri}^{\cal P}}\bar{l}^e E^{\cP}_{\rm sq}(\cN^e)
\ee
for any partition ${\cal P}$ of the set ${\cal V}$, where $\rho_0^{\cal V}$ is the initial state of the whole system, $\langle f({{\bm k}_i}) \rangle_{{\bm k}_i}:=\sum_{{\bm k}_i} p({{\bm k}_i}) f({{\bm k}_i})$. Further, $\bar{l}^e=\sum_{i=1}^{l}\sum_{{\bm k}_{i-1}}p({\bm k}_{i-1})\delta_{e,e_{{\bm k}_{i-1}}}$ is the average number of uses of channel $\cN^e$ in the protocol.
\end{theorem}
The proof of Theorem \ref{lemma:main} is given in Appendix \ref{app:A}. 

Let us now assume that the quantum state of families ${\cal S}_1,\ldots,{\cal S}_m$ after $l$ rounds is $\epsilon$-close to our target state, i.e.,
\be
\| \r_{{\bm k}_l}^{{\cal T}_m}-\tau^{{\cal T}_m}_{{\bm d}_{{\bm k}_l}}\|_1\le\e \label{err}
\ee
with an error parameter $\epsilon>0$, where $\r^{{\cal T}_m}_{{\bm k}_l}={\rm Tr}_{{\cal V} \setminus {\cal T}_m} (\r_{{\bm k}_l}^{\cal V})$.
Theorem \ref{lemma:main} can provide us with an upper bound on the achievable squashed entanglement of $\tau_{{\bm d}_{k_l}}^{{\cal T}_m}$. The question now is how to choose partition $\cP$. It holds for any partition $\cP$ that
\be
E_{\rm sq}^{\cP}(\tau_{\bm d}^{{\cal T}_m})=\sum_{j=1}^m E_{\rm sq}^{\cP}(\gamma_{d^{(j)}}^{{\cal S}_j}) \ge \sum_{j=1}^mn^{\cS_j|\cP}\log d^{(j)},
\ee
where $n^{\cS_j|\cP}$ is the number of parts the partition ${\cal P}$ nontrivially divides ${\cS_j}$ into. This is formally defined by
\be\label{def:k}
n^{\cS_j|\cP}=\begin{cases} 
0 & (|\{ l \in \{1,\ldots,k\}|{\cal P}_l \cap {\cal S}_j \neq \emptyset \}| < 2)\\ 
|\{ l \in \{1,\ldots,k\}|{\cal P}_l \cap {\cal S}_j \neq \emptyset \}|& (|\{ l \in \{1,\ldots,k\}|{\cal P}_l \cap {\cal S}_j \neq \emptyset \}| \ge 2)
\end{cases} 
\ee
for ${\cal P}=\cP_1:\cdots:\cP_k$. Hence, the squashed entanglement w.r.t. any partition ${\cal P}$ with $n^{{\cal S}_j|{\cal P}}\neq 0$ can provide an upper bound on the size of distillable secret key or GHZ state of family $S_j$ in the protocol. 
\begin{theorem}\label{EsqBoundMark}
If an adaptive protocol, having started from the initial state $\r^{{\cal V}}_0$, after $l$ rounds results in a general target state $\tau^{{\cal T}_m}_{{\bm d}_{{\bm k}_l}}$ of form (\ref{target}) within an error $\epsilon>0$, it holds for any partition $\cP$ that
\be\label{eq:EsqBound}
\sum_{j=1}^mn^{\cS_j|\cP}\langle \log d^{(j)}_{{\bm k}_l}\rangle_{{\bm k}_l} \le \frac{1}{1-b\e}\left(E_{\rm sq}^{\cP}(\r^{{\cal V}}_0)+\sum_{ e\in {\cal E}\setminus {\cal E}_{\rm tri}^{\cal P}   }\bar{l}^e  E^{\cP}_{\rm{\rm sq} }\left(\cN^{e}\right)+g(\e)\right),
\ee
where $n^{\cS_j|\cP}$ is defined in (\ref{def:k}), $b\in\ZZ^+$ and $g(\e)\to0$ as $\e\to0$.
\end{theorem}

This theorem is a generalisation of the upper bounds \cite{azuma2016fundamental} for multi-pair bipartite entanglement distribution protocols. One can minimise the r.h.s of (\ref{eq:EsqBound}) over partitions $\cP$, with constraints on $n^{\cS_j|\cP}$ depending on the user scenario. An example will be given below in Corollary \ref{GHZBound}. In order to obtain a tighter bound one could also use a set of inequalities of form (\ref{eq:EsqBound}) w.r.t. different partitions. Before proving Theorem \ref{EsqBoundMark}, we need to show that the the multipartite squashed entanglement is strongly superadditive, as has been shown in the bipartite case \cite{christandl2006structure}. 
\begin{lemma}\label{EsqSSA}
For a quantum state $\r^{A_1A_1'\cdots A_mA_m'}$ with marginals $\r^{A_1\cdots A_m}$ and $\r^{A_1'\cdots A_m'}$, it holds
\be
E_{\rm sq}^{A_1A_1':\cdots:A_mA_m'}\left(\r^{A_1A_1'\cdots A_mA_m'}\right)\ge E_{\rm sq}^{A_1:\cdots:A_m}\left(\r^{A_1..A_m}\right)+E_{\rm sq}^{A_1':\cdots:A_m'}\left(\r^{A_1'\cdots A_m'}\right).
\ee
\end{lemma}
\proof
Let $\r^{A_1A_1'\cdots A_mA_m'E}$ be an extension of $\r^{A_1A_1'\cdots A_mA_m'}$. By Corollary 1 of \cite{yang2009squashed} it holds
\be
I(A_1A_1':\cdots:A_mA_m'|E)\ge I(A_1:\cdots:A_m|EA_1'\cdots A_m')+I(A_1':\cdots:A'_m|E).
\ee
As the squashed entanglement is the infimum over all extensions, the Lemma follows.\newline
\qed
We are now ready to prove Theorem \ref{EsqBoundMark}.\newline

\proofof {\bf Theorem \ref{EsqBoundMark}.}
Let $\cP$ be a partition.
As $\| \r_{{\bm k}_l}^{{\cal T}_m}-\tau_{{\bm d}_{{\bm k}_l}}^{{\cal T}_m} \|_1\le\e$, it holds $\| \r_{{\bm k}_l}^{\cS_j}-\gamma^{\cS_j}_{d^{(j)}_{{\bm k}_l}} \|_1\le\e$, and thus $F(\r_{{\bm k}_l}^{\cS_j},\gamma^{\cS_j}_{d^{(j)}_{{\bm k}_l}})\ge1-\e$ for every $j\in\{1,\ldots,m\}$. By \cite{wilde2016squashed}, for any $j$ with $n^{{\cal S}_j|{\cal {P}}} \neq 0$, this implies
\be
n^{\cS_j|\cP}\log d^{(j)}_{{\bm k}_l}\le E_{\rm sq}^{\cP}(\r_{{\bm k}_l}^{\cS_j})+n^{\cS_j|\cP} b_j\e\log d^{(j)}+{f}_j(\e),
\ee
where $b_j\in \ZZ^+$ and ${f}_j(\e)\to0$ as $\e\to0$. This inequality trivially holds for $j$ with $n^{{\cal S}_j|{\cal {P}}} = 0$, by defining $b_j=f_j(\epsilon)=0$. 
Hence, for any $j$, we have
\be
n^{\cS_j|\cP}\log d^{(j)}_{{\bm k}_l}\le\frac{1}{1-b\e}\left( E_{\rm sq}^{\cP}(\r_{{\bm k}_l}^{\cS_j})+{f}_j(\e)\right),
\ee
with $b=\max_j \{b_j\}$. By application of Lemma \ref{EsqSSA} and the reduction property (\ref{reduction}) it holds
\be
\sum_{j}n^{\cS_j|\cP}\log d^{(j)}_{{\bm k}_l}\le\frac{1}{1-b\e}\left(  E_{\rm sq}^{\cP}(\r_{{\bm k}_l}^{{\cal T}_m})+g(\epsilon)\right),
\ee
where we have defined $g(\epsilon)=\sum_{j=1}^m{f}_j(\e)$. As tracing out subsystems cannot increase the squashed entanglement, application of Theorem \ref{lemma:main} finishes the proof.\newline
\qed
If the goal of the protocol is to only distribute a single private state $\gamma_{d_{{\bm k}_l}}^{{\cal S}}$ among one family ${\cal S}$ of nodes, that is, if the target state is $\tau_{{\bm d}_{{\bm k}_l}}^{{\cal T}_m}$ with $m=1$, ${\cal S}_1={\cal S}$ and $d^{(1)}_{{\bm k }_l}=d_{{\bm k }_l}$, then Theorem~\ref{EsqBoundMark} can be reduced into a simpler form.

\begin{corollary}\label{GHZBound}
If an adaptive protocol starting from initial state $\r_0^{\cal V}$, after $l$ rounds, provides finite dimensional $\gamma_{d_{\bm{k}_l}}^{\cS}$ within an error $\epsilon>0$, 
\be
 \left\langle\log d_{{{\bm k}_l}}\right\rangle_{{\bm k}_l}\le\min_{\cP:n^{{\cal S}|{\cal P}} \neq0}\frac{1}{n^{{\cal S}|{\cal P}}(1-b\e)}\left(E_{\rm sq}^{\cP}(\r_0^{{\cal V}})+\sum_{ e\in {\cal E}\setminus {\cal E}_{\rm tri}^{\cal P} }\bar{l}^e  E^{\cP}_{\rm sq}\left(\cN^{e}\right)+g(\e)\right)
\ee
holds, where the minimisation is over all possible partitions $\cP$ with $n^{{\cal S}|{\cal P}} \neq0$. Further, $b\in\ZZ^+$ and $g(\e)\to0$ as $\e\to0$.
\end{corollary}
In the case of only two clients, Corollary \ref{GHZBound} reduces to the bound provided in \cite{azuma2016fundamental}.

\section{How to obtain lower bounds?}\label{sec:lower}
A natural question arising now is whether there exist lower bounds on the rates discussed above. In the case where a single pair of clients wish to establish a bipartite key or Bell states by using a quantum network composed only of bipartite channels, this question has been addressed in \cite{azuma2016aggregating}. Their lower bound is achieved by means of a so-called \emph{aggregated quantum repeater protocol}. This protocol begins by distribution of a number of Bell states via each channel. The resulting network of Bell states is then used to distribute maximal entanglement between Alice and Bob by means of entanglement swapping. The network of Bell states can be described by an undirected graph. The amount of maximal entanglement obtainable in this way depends on the number of edge-disjoint paths between Alice and Bob. By Menger's theorem \cite{menger1927allgemeinen}, this number is equal to the minimum number of edges in an Alice-Bob cut. In the case of many clients and broadcast networks, the situation becomes more involved. In the following, we briefly describe how lower bounds can be obtained. We will present a rigorous graph theoretic derivation of the bounds in a future work.

\begin{figure}

	\centering
	(a)\includegraphics[width=0.4\textwidth]{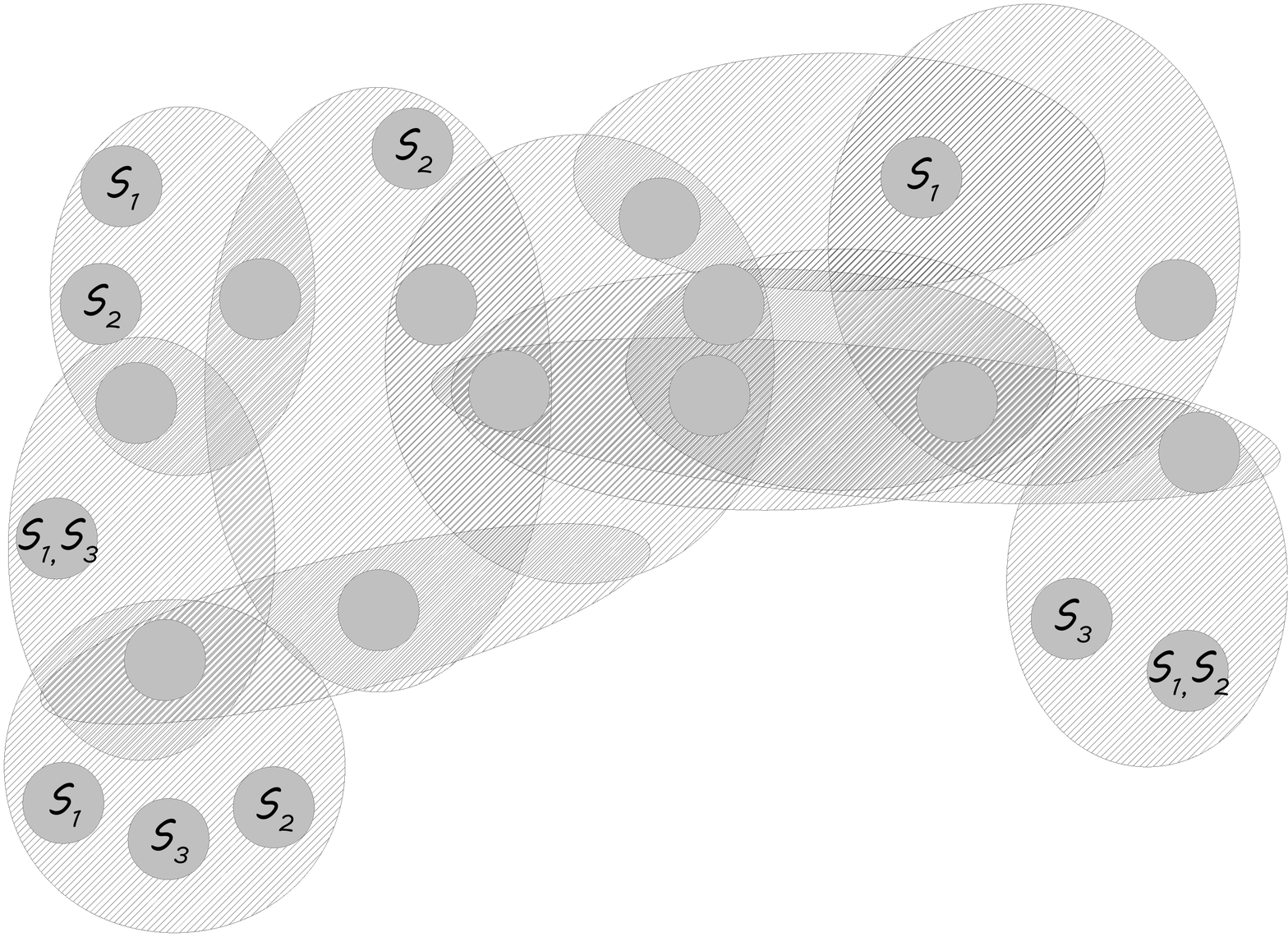}
	(b)\includegraphics[width=0.4\textwidth]{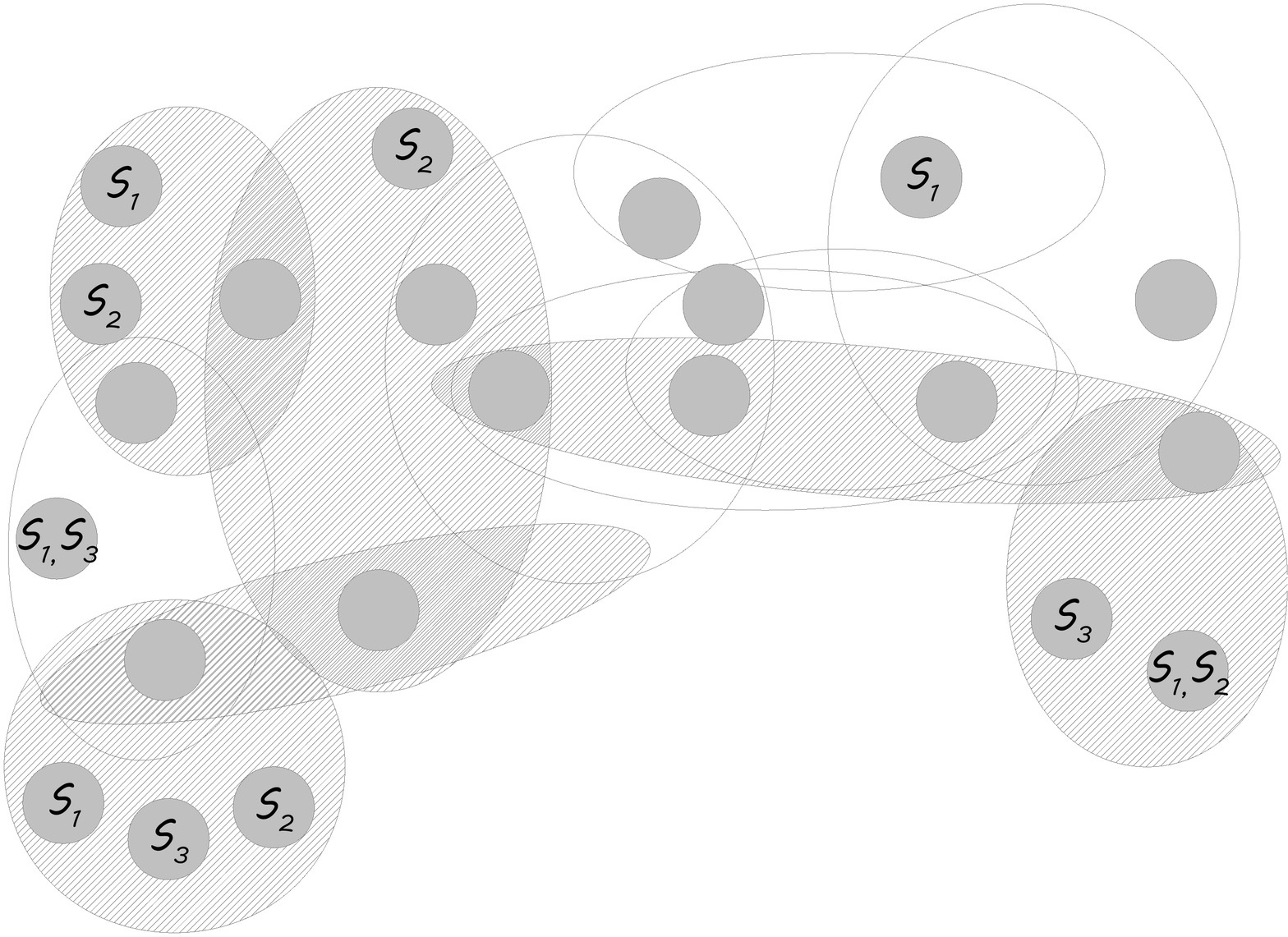}
	\caption{(a) The network from figure \ref{fig1} transformed into a network of GHZ states described by an undirected hypergraph $H^\text{GHZ}$. Here we have assumed that each channel has produced exactly one GHZ state. (b) Example of a Steiner hypertree spanning $\cS_2$.}
\label{fig3}
\end{figure}

As in the bipartite case, we can compose an aggregated repeater protocol even for such a general case. In the protocol, we begin by using each broadcast channel to distribute a number of qubit GHZ states. This results in a network of GHZ states, which can be associated with an undirected hypergraph $H^\text{GHZ}$. See also figure \ref{fig3}a. The GHZ network can be transformed into the desired target state by a generalised version of entanglement swapping \cite{wallnofer20162d}. Namely, it has been shown that an $n$-partite GHZ state among parties $A_1\cdots A_n$ and an $m$-partite GHZ state among parties $A'_1\cdots A'_m$ can be transformed to a single $(n+m-1)$-partite GHZ state by means of a projection onto a pair $A_iA'_j$ of parties, followed by classical communication and local Pauli corrections. In addition, it is possible to transform an $n$-partite GHZ state into an $(n-1)$-partite GHZ state by a measurement in the $\s_x$ eigenbasis and a local correction with unitary $\s_z$ depending on the measurement outcome. These techniques can be used to convert $H^\text{GHZ}$ into the desired target state.

The main challenge is to compute the achievable dimension of the target state by a suitable generalisation of Menger's theorem. If the goal is to only distribute entanglement between two clients, Alice and Bob, via a network consisting of broadcast channels, we can apply a generalisation of Menger's theorem to paths consisting of hyperedges as provided in \cite{kiraly2003edge}. In the case of many clients, however, we need a generalisation of the concept of paths. Assuming we want to distribute GHZ entanglement among a family $\cS$ of clients, we need to obtain the number of edge-disjoint \emph{Steiner trees} in $H^\text{GHZ}$ that span $\cS$:  A Steiner tree in a (hyper)graph spanning a set of vertices $\cS$ is defined as an acyclic sub-(hyper)graph connecting all vertices in $\cS$ \cite{brazil2015steiner}. See also figure \ref{fig3}b. Each Steiner tree in $H^\text{GHZ}$ spanning $\cS$ can be transformed into one qubit GHZ state among the clients in $\cS$ by means of the generalised entanglement swapping protocol described above. The problem of finding the number of edge-disjoint Steiner trees is referred to \emph{Steiner tree packing}. Even in the case of graphs, Steiner tree packing has been shown to be an NP complete problem \cite{kaski2004packing}. There are, however, polynomial algorithms, that can provide us with lower bounds on the number of edge-disjoint Steiner trees in a graph \cite{jain2003packing,lau2004approximate}. The result of \cite{lau2004approximate} connects the maximum number of edge-disjoint Steiner trees spanning a set $\cS$ of vertices in a graph to the minimum Steiner cut w.r.t. $\cS$. By a minimum Steiner cut w.r.t. $\cS$  we mean the smallest set of edges whose removal disconnects some pair of vertices in $\cS$. Concretely, it is shown that if the minimum Steiner cut w.r.t. $\cS$ contains $26k$ edges, the graph will have, at least, $k$ edge-disjoint Steiner trees spanning $S$. The number of edges in the minimum Steiner cut can be computed in polynomial time. This can provide us with a lower bound on the rate at which GHZ states can be distributed in a network consisting only of single-sender-receiver channels. In the case of broadcast channels we need to solve the Steiner tree packing problem for hypergraphs, which we leave for further research. If the goal is to distribute GHZ states among different families of clients in parallel, a more complex graph theoretic analysis, combining Steiner trees with concepts like multi-commodity flows \cite{korte2008multicommodity,chalermsook2012approximation}, will be necessary. We leave this question open for further research, as well.

\section{Conclusion and Outlook}\label{sec:concl}
We have provided a general framework to derive upper bounds on the rates at which bi- and multipartite entanglement can be distributed in various multi-user scenarios in a network consisting of quantum broadcast channels. Our theorem \ref{lemma:main}, upper bounds the multipartite squashed entanglement of the achievable target state w.r.t. arbitrary partitions. By choosing appropriate partitions, this can provide upper bounds on the rates at which Bell and GHZ states can be distributed in parallel between arbitrary families of users, as represented by theorem \ref{EsqBoundMark}. Upper bounds can also be applied to distribution of multipartite private states. The bound in theorem \ref{EsqBoundMark} is obtained by summing over the maximal squashed entanglement distributable in every use of a broadcast channel across the partition $\cP$. While we have concentrated on the squashed entanglement, it has been shown that the relative entropy of entanglement and various Renyi generalisations can provide upper bounds on key rates \cite{2015arXiv151008863P,pirandola2016optimal,takeoka2016unconstrained,wilde2016converse,christandl2016relative}. We leave it as an open question whether these bounds can also be generalised to the case of quantum broadcast networks. Another open question is whether bounds can also be obtained for a network of multiple access channels. A special case of multiple access channels are noisy non-local gates, as used in a quantum computer. Hence, if we generalise theorem \ref{EsqBoundMark} to include multiple access channels, it might be possible to apply such a bound to a quantum computer, which could be described as a network of noisy gates. Another future direction of research will be to optimise the protocols w.r.t. the amount of experimental resources or the time needed. For example, one could minimise the number of repeater stations needed or concentrate on networks with only one-way classical communication. We have also discussed how lower bounds can be achieved by means of an aggregated quantum repeater protocol. Such a protocol uses the broadcast channels to create a network of GHZ states, which are then connected to form the target state by means of LOCC. The obtainable dimensions of the target state can be obtained by a graph theoretic analysis of the corresponding hypergraph. In the case of a single GHZ state,  this is done by identifying the number of edge-disjoint Steiner trees spanning the set of parties involved.

\section*{Acknowledgements}
Among others, we would like to thank Go Kato, Bill Munro, Kae Nemoto, Simone Severini, Mark Wilde, Karol Horodecki, David Elkouss, Masahiro Takeoka, Hayata Yamasaki, Stefano Pirandola and Harumichi Nishimura for valuable discussions. K.A. thanks support from the ImPACT Program of Council for Science, 
Technology and Innovation (Cabinet Office, Government of Japan).
\newpage
\begin{appendix}
\section{Proof of Theorem \ref{lemma:main}}\label{app:A}
In order to prove Theorem~\ref{lemma:main}, we need a variation of Lemma 6 in \cite{seshadreesan2015bounds}:
\begin{lemma}\label{lemma:subadd}
Let $n\le m$ and $\ket{\psi}^{SP_1\cdots P_mQ_1\cdots Q_nE_1E_2}$ be a pure state. Then it holds
\begin{align}
&E_{\rm sq}^{S:P_1Q_1:\cdots :P_nQ_n:P_{n+1}:\cdots :P_m}(\pro{\psi}) \nonumber \\
&\le E_{\rm sq}^{SQ_1\cdots Q_nE_2:P_1:\cdots :P_m}(\pro{\psi})+E_{\rm sq}^{SP_1\cdots P_mE_1:Q_1:\cdots :Q_n}(\pro{\psi}).
\end{align}
\end{lemma}
The proof of Lemma \ref{lemma:subadd} is similar to the one in \cite{seshadreesan2015bounds}. The difference is that the squashed entanglement for different numbers of parties is involved. For completeness, we have included it here:

\proofof\textbf{Lemma \ref{lemma:subadd}.}
Let $\ket{\psi}^{SP_1\cdots P_mQ_1\cdots Q_nE_1 E_2}$ be a pure state and let $\Lambda^1:{E_1\to E'_1}$ and $\Lambda^2:{E_2\to E'_2}$ be local squashing channels in the sense of \cite{seshadreesan2015bounds}. We define 
\begin{align*}
&\tau^{SP_1\cdots P_mQ_1\cdots Q_nE'_1E_2}=\id\te\Lambda^1(\pro{\psi}),\\
&\s^{SP_1\cdots P_mQ_1\cdots Q_nE_1E'_2}=\id\te\Lambda^2(\pro{\psi}),\\
&\omega^{SP_1\cdots P_mQ_1\cdots Q_nE'_1E'_2}=\id\te\Lambda^1\te \Lambda^2(\pro{\psi}).
\end{align*}
Let $\ket{\omega}^{SP_1\cdots P_mQ_1\cdots Q_nE'_1E'_2R}$ be a purification of $\omega$. It holds
\begin{align*}
&E_{\rm sq}^{S:P_1Q_1:\cdots :P_nQ_n:P_{n+1}:\cdots :P_m}(\pro{\psi})\\
&\le I(S:P_1Q_1:\cdots :P_nQ_n:P_{n+1}:\cdots :P_m|E'_1E'_2)_\omega\\
&= H(SE'_1E'_2)_\omega+\sum_{i=1}^nH(P_iQ_i|E'_1E'_2)_\omega+\sum_{i=n+1}^mH(P_i|E'_1E'_2)_\omega-H(SP_1Q_1\cdots P_nQ_nP_{n+1}\cdots P_mE'_1E'_2)_\omega\\
&= \sum_{i=1}^nH(P_iQ_i|E'_1E'_2)_\omega+\sum_{i=n+1}^mH(P_i|E'_1E'_2)_\omega-H(P_1Q_1\cdots P_nQ_nP_{n+1}\cdots P_m|E'_1E'_2S)_\omega\\
&=\sum_{i=1}^nH(P_iQ_i|E'_1E'_2)_{\pro{\omega}}+\sum_{i=n+1}^mH(P_i|E'_1E'_2)_{\pro{\omega}}+H(P_1Q_1\cdots P_nQ_nP_{n+1}\cdots P_m|R)_{\pro{\omega}}\\
&\le \sum_{i=1}^mH(P_i|E'_1)_{\pro{\omega}}+ \sum_{i=1}^nH(Q_i|E'_2)_{\pro{\omega}}+H(P_1\cdots P_m|R)_{\pro{\omega}}+H(Q_1\cdots Q_n|R)_{\pro{\omega}}\\
&= \sum_{i=1}^mH(P_i|E'_1)_{\omega}-H(P_1\cdots P_m|SQ_1\cdots Q_nE'_1E'_2)_{\omega}+ \sum_{i=1}^nH(Q_i|E'_2)_{{\omega}}-H(Q_1\cdots Q_n|SP_1\cdots P_mE'_1E'_2)_{\omega}\\
&=I(SQ_1\cdots Q_nE'_2:P_1:\cdots :P_m|E'_1)_\omega+I(SP_1\cdots P_mE'_1:Q_1:\cdots :Q_n|E'_2)_\omega\\
&\le I(SQ_1\cdots Q_nE_2:P_1:\cdots :P_m|E'_1)_\tau+I(SP_1\cdots P_mE_1:Q_1:\cdots :Q_n|E'_2)_\s,
\end{align*}
where we have used the strong subadditivity of the von Neumann entropy \cite{lieb2002proof} in the second inequality and the data processing inequality in the last inequality. Note that this holds for all local squashing channels $\Lambda^1:{E_1\to E'_1}$ and $\Lambda^2:{E_2\to E'_2}$. In particular
\begin{align*}
&E_{\rm sq}^{S:P_1Q_1:\cdots :P_nQ_n:P_{n+1}:\cdots :P_m}(\pro{\psi})\\
&\le\inf_{\Lambda^1:{E_1\to E'_1}}I(SQ_1\cdots Q_nE_2:P_1:\cdots :P_m|E'_1)_{\Lambda^1(\pro{\psi})}+\inf_{\Lambda^2:{E_2\to E'_2}}I(SP_1\cdots P_mE_1:Q_1:\cdots :Q_n|E'_2)_{\Lambda^2(\pro{\psi})}\\
&=E_{\rm sq}^{SQ_1\cdots Q_nE_2:P_1:\cdots :P_m}(\pro{\psi})+E_{\rm sq}^{SP_1\cdots P_mE_1:Q_1:\cdots :Q_n}(\pro{\psi}),
\end{align*}
finishing the proof. \newline
\qed
Before proving Theorem \ref{lemma:main}, let us introduce the following notation: For a set $\cal V'\subset V$ of vertices, we define the Hilbert space of the quantum system held by the vertices in $\cal V'$ at step $i$ of the protocol as $\cH^i(\cal V')$.\newline

\proofof\textbf{Theorem \ref{lemma:main}.}
Let $\cP$ be some partition of the vertices into $k$ disjoint classes $\cP_1,\cP_2,\ldots,\cP_k$. Consider a generic round $i$ of an adaptive LOCC protocol as described in section \ref{sec:Notation}. Let ${\bm k}_{i-1}$ be the vector of outcomes of the previous rounds and let the state of the entire system be given by $\r_{{\bm k}_{i-1}}^{\cal V}$ on $\cH^{i-1}(\cal V)$. Depending on the protocol and the previous outcomes ${\bm k}_{i-1}$ either a broadcast channel is used, followed by LOCC, or an LOCC operation is performed without use of any channel. If we use a broadcast channel $\cN^{e_{{\bm k}_{i-1}}}$ that crosses\footnote{We say a broadcast channel $\cN^{e}$ crosses a partition, if at least one element of $h(e)$ is in a class different from the one $t(e)$ belongs to.} partition $\cP$, we write ${\bm k}_{i-1}\in K^{\cP}$. 
 After the possible use of the broadcast channel an arbitrary LOCC protocol is applied, resulting in  $\r_{{\bm k}_i}^{\cal V}$ on $\cH^{i}(\cal V)$ with probability $p({k_i|{\bm k}_{i-1}})$. 

In the case where no channel crossing $\cP$ is used, i.e. ${\bm k}_{i-1}\notin K^{\cP}$, it holds by the LOCC monotonicity of the squashed entanglement that
 \be
\sum_{k_i}p({k_i|{\bm k}_{i-1}}) E_{\rm sq}^{\cH^i(\cP_1):\cdots : \cH^i(\cP_k)}(\r_{{\bm k}_i}^{\cal V})\le E_{\rm sq}^{\cH^{i-1}(\cP_1):\cdots : \cH^{i-1}(\cP_k)}(\r_{{\bm k}_{i-1}}^{\cal V}).
\ee

Let us now consider the case where a broadcast channel $\cN^{e_{{\bm k}_{i-1}}}$ crossing $\cP$ is used, i.e. ${\bm k}_{i-1}\in K^{\cP}$. The input of $\cN^{e_{{\bm k}_{i-1}}}$ consists of a subsystem $\cH_x^{{\bm k}_{i-1}}$ of the quantum system of the unique vertex in $t({e_{{\bm k}_{i-1}}})$. The output systems $\cH_{y_1}^{{\bm k}_{i-1}},\ldots,\cH_{y_r}^{{\bm k}_{i-1}}$ become part of the quantum systems of the respective vertices in $h({e_{{\bm k}_{i-1}}})$. Let $j_0\in\{1,\ldots,k\}$ be such that $t({e_{{\bm k}_{i-1}}})\subset\cP_{j_0}$ and let $s\ge1$ be the smallest number such that $h({e_{{\bm k}_{i-1}}})\subset\bigcup_{v=0}^s\cP_{j_v}$ for distinct $j_1,\ldots,j_s\in\{1,\ldots,k\}$. Note that $j_1,\ldots,j_s$ are different from $j_0$. For all $j \in \{  1,\ldots, s \}$, ${\cal H}_{Y_{j_l}}^{{\bm k}_i-1}$ denotes the combined system of all the output systems that become part of ${\cal H}^i( {\cal P}_{j_l})$.
The quantum system of the class $\cP_{j_0}$ after using the channel, containing the sender, is given by $\cH^{i}(\cP_{j_0})$. In general, some of the vertices in $h(e_{{\bm k}_{i-1}})$ might also be in $\cP_{j_0}$. In this case we assume that $\cH^{i}(\cP_{j_0})$ also contains the corresponding output systems. Then we now apply Lemma \ref{lemma:subadd} in addition to the LOCC monotonicity of the squashed entanglement and obtain\begin{align*}
&\sum_{k_i}p({k_i|{\bm k}_{i-1}}) E_{\rm sq}^{\cH^i(\cP_1):\cdots : \cH^i(\cP_k)}(\r_{{\bm k}_i}^{\cal V})\\
&\le E_{\rm sq}^{\cH^{i-1}(\cP_1):\cdots : \cH^{i}(\cP_{j_0}): \cH^{i-1}(\cP_{j_1}) \cH_{Y_{j_1}}^{{\bm k}_{i-1}}:\cdots : \cH^{i-1}(\cP_{j_s}) \cH_{Y_{j_s}}^{{\bm k}_{i-1}}:\cdots : \cH^{i-1}(\cP_k)}\left(\cU^{{\bm k}_{i-1}}\pro{\r^{\cal V}_{{\bm k}_{i-1}}}\cU^{{\bm k}_{i-1}\dagger}\right)\\
&\le E_{\rm sq}^{\cH^{i-1}(\cP_1):\cdots : \cH^{i}(\cP_{j_0}) \cH_{Y_{j_1}}^{{\bm k}_{i-1}}\cdots  \cH_{Y_{j_s}}^{{\bm k}_{i-1}}\cH_{E}^{{\bm k}_{i-1}}: \cH^{i-1}(\cP_{j_1}):\cdots : \cH^{i-1}(\cP_{j_s}):\cdots : \cH^{i-1}(\cP_k)}\left(\cU^{{\bm k}_{i-1}}\pro{\r^{\cal V}_{{\bm k}_{i-1}}}\cU^{{\bm k}_{i-1}\dagger}\right)\\
&+E_{\rm sq}^{\cH^{i-1}(\cP_1)\cdots \cH^{i}(\cP_{j_0})\cH^{i-1}(\cP_{j_1})\cdots \cH^{i-1}(\cP_{j_s})\cdots  \cH^{i-1}(\cP_k)\cH_{R}^{{\bm k}_{i-1}}: \cH_{Y_{j_1}}^{{\bm k}_{i-1}}:\cdots : \cH_{Y_{j_s}}^{{\bm k}_{i-1}}}\left(\cU^{{\bm k}_{i-1}}\pro{\r^{\cal V}_{{\bm k}_{i-1}}}\cU^{{\bm k}_{i-1}\dagger}\right)\\
&\le E_{\rm sq}^{\cH^{i-1}(\cP_1):\cdots : \cH^{i-1}(\cP_k)}(\r_{{\bm k}_{i-1}}^{\cal V})+E_{\rm sq}^{\cP}(\cN^{e_{{\bm k}_{i-1}}}),
\end{align*}
where $\cU^{{\bm k}_{i-1}}:\cH_x^{{\bm k}_{i-1}} \to\cH_{y_1}^{{\bm k}_{i-1}}\cdots \cH_{y_r}^{{\bm k}_{i-1}}\cH_{E}^{{\bm k}_{i-1}}$ is the isometric extension of $\cN^{e_{{\bm k}_{i-1}}}$ and $\ket{\r^{\cal V}_{{\bm k}_{i-1}}}$ on an extended system $\cH^{i-1}({\cal V})\cH_{R}^{{\bm k}_{i-1}}$ is a purification of $\r_{{\bm k}_{i-1}}^{\cal V}$. Note that this is a recursive relation. If we now start from $i=l$ and backtrack to $i=1$, recursively applying the above relation, we obtain

\begin{align}
&\sum_{{\bm k}_l}p({{\bm k}_l})E_{\rm sq}^{\cH^l(\cP_1):\cdots : \cH^l(\cP_k)}(\r_{{\bm k}_l}^{\cal V})\\
&=\sum_{{\bm k}_{l-1}}p({{\bm k}_{l-1}})\sum_{k_l}p(k_l|{{\bm k}_{l-1}})E_{\rm sq}^{\cH^l(\cP_1):\cdots : \cH^l(\cP_k)}(\r_{{\bm k}_l}^{\cal V})\\
&\le\sum_{{\bm k}_{l-1}}p({{\bm k}_{l-1}})E_{\rm sq}^{\cH^{l-1}(\cP_1):\cdots : \cH^{l-1}(\cP_k)}(\r_{{\bm k}_{l-1}}^{\cal V})+\sum_{{\bm k}_{l-1}\in K^{\cP}}p({{\bm k}_{l-1}})E_{\rm sq}^{\cP}(\cN^{e_{{\bm k}_{l-1}}})\\\
&\le\cdots \\
&\le\sum_{k_1}p(k_1)E_{\rm sq}^{\cH^1(\cP_1):\cdots : \cH^1(\cP_k)}(\r_{k_1}^{\cal V})+\sum_{i=2}^{l}\sum_{{\bm k}_{i-1}\in K^{\cP}}p({\bm k}_{i-1})E^{\cP}_{\rm sq}(\cN^{e_{{\bm k}_{i-1}}})\\
&\le E_{\rm sq}^{\cH^0(\cP_1):\cdots : \cH^0(\cP_k)}(\r_0^{\cal V})+\sum_{i=1}^{l}\sum_{{\bm k}_{i-1}\in K^{\cP}}p({\bm k}_{i-1})E^{\cP}_{\rm sq}(\cN^{e_{{\bm k}_{i-1}}})\\
&=E_{\rm sq}^{ \cH^0(\cP_1):\cdots : \cH^0(\cP_k)}(\r_0^{\cal V})+\sum_{e\in \cE\setminus\cE^{\cP}_\text{tri}}\bar{l}^eE^{\cP}_{\rm sq}(\cN^{e}),
\end{align}
where we have defined $\bar{l}^e=\sum_{i=1}^{l}\sum_{{\bm k}_{i-1}}p({\bm k}_{i-1})\delta_{e,e_{{\bm k}_{i-1}}}$, which is the average number of uses of channel $\cN^e$ in the protocol.
\qed

\end{appendix}

\bibliographystyle{unsrt}
\bibliography{Network.bib}
\end{document}